\documentclass[preprint,showpacs,preprintnumbers,nofootinbib]{revtex4}

\usepackage{graphicx}
\usepackage{dcolumn}
\usepackage{bm}

\newcommand{\bea}{\begin{eqnarray}}
\newcommand{\eea}{\end{eqnarray}}
\newcommand{\be}{\begin{equation}}
\newcommand{\ee}{\end{equation}}
\newcommand{\pp}{\psi^{\prime}}
\newcommand{\xp}{\xi^{\prime}}
\newcommand{\cp}{\chi^{\prime}}
\newcommand{\fp}{\phi^{\prime}}
\newcommand{\vp}{\varphi^{\prime}}
\newcommand{\lpp}{\lambda^{\prime \prime}}
\newcommand{\npp}{\nu^{\prime \prime}}
\newcommand{\np}{\nu^{\prime}}
\newcommand{\lp}{\lambda^{\prime}}
\newcommand{\lps}{\lambda^{\prime 2}}
\newcommand{\nps}{\nu^{\prime 2}}
\newcommand{\gn}{8 \pi M_{p}^{-2}}
\newcommand{\ppp}{\psi^{\prime\prime}}
\newcommand{\xpp}{\xi^{\prime\prime}}
\newcommand{\cpp}{\chi^{\prime\prime}}
\newcommand{\Dpp}{\Delta^{\prime\prime}}
\newcommand{\Tpp}{\Theta^{\prime\prime}}
\newcommand{\Dp}{\Delta^{\prime}}
\newcommand{\Tp}{\Theta^{\prime}}

\newcommand{\vpp}{\varphi^{\prime\prime}}

\newcommand{\ch}{{\cal{H}}}
\begin{document}

\preprint{Brown-HET-1394}

\title{UV Perturbations in Brane Gas Cosmology}

\author{Scott Watson}
 \email{watson@het.brown.edu}%
\affiliation{Department of Physics, Brown University, Providence, RI.}

\date{\today}

\begin{abstract}
We consider the effect of the ultraviolet (UV) or short wavelength modes
on the background of Brane Gas Cosmology.
We find that the string matter sources are negligible
in the UV and that the evolution is given primarily by the dilaton
perturbation.  We also find that the linear perturbations are
well behaved and the predictions of Brane Gas Cosmology are
robust against the introduction of linear perturbations. In
particular, we find that the stabilization of the extra dimensions
(moduli) remains valid in the presence of dilaton and string
perturbations.
\end{abstract}

\pacs{Valid PACS appear here}
\maketitle

\section{Introduction}

Understanding the behavior of strings in a time dependent
background has been a subject of much interest and has been
pursued in a number of differing ways. One scenario, known as
Brane Gas Cosmology (BGC), is devoted to understanding the effect
that string and brane gases could have on a dilaton-gravity
background in the early Universe
\cite{bv,vafa,bgc,isotropization,stable,extended}. In \cite{bv},
it was suggested that the energy associated with the winding of
strings around the compact dimensions would produce a confining
potential for the scale factor and halt the cosmological
expansion\footnote{This was later shown quantitatively in
\cite{vafa}.}. The analysis of BGC was initially performed under the
assumption of a homogeneous and isotropic cosmology.
The results were recently extended to the case of anisotropic
cosmology in \cite{isotropization}. There, it was shown that
string gases can give rise to three dimensions growing large and
isotropic due to string annihilation while the other six
dimensions remain confined. In \cite{stable} it was shown that by
considering both momentum and winding modes of strings, the six
confined dimensions can be stabilized at the self-dual radius,
where the energy of the string gas is minimal. This result
demonstrated that, in BGC, the volume moduli of the extra
dimensions can be stabilized in a natural and intuitive way.

In recent work \cite{perturbations}, we considered the effect of string inhomogeneities
and dilaton fluctuations on BGC.  The string sources of BGC
are usually represented by a perfect fluid with homogeneous energy and
pressure densities given by the mass spectrum of the strings
(see e.g. \cite{bgc,stable,subodh}).
One may worry that inhomogeneities
of string sources (in particular strings winding around the confined
dimensions) as a function of the unconfined spatial directions could
lead to serious instabilities which could ruin the main
successes of BGC, namely the prediction that three directions become
large leaving the other six confined uniformly as a function of the
coordinates of the large spatial sections.
In \cite{perturbations}, we found that at the linear
level BGC is robust with respect to long wavelength perturbations. In that paper it was found
that at late times the inhomogeneities are subleading compared to the evolution of the background.
In this paper
we will extend our considerations to the ultraviolet or small wavelength perturbations.
Our expectation was that on small wavelengths, the motion of the strings would smear
out potential instabilities in a way analogous to how the motion of light particles
(``free-streaming'') leads to a decay of short wavelength fluctuations
in standard cosmology (see e.g. \cite{Peebles} for a review).
However, we will find that the string matter perturbations are
actually sub-leading in the evolution and the dilaton perturbation
is the primary driving force of instability.

For reference, in Section 2 and 3 we present the background solution and
perturbed equations as found in \cite{perturbations}.  The crucial new results
appear in Section 4, where we derive the perturbation equations for
the UV modes and then solve for their late time behavior.  The full equations
are presented in the Appendix. We conclude with a discussion of our findings and
future prospects in Section 5.

\section{Background Solution}

Our starting point is the low energy effective action for the bulk space-time
with string matter sources \cite{vafa},
\be
\label{action}
S=\frac{1}{4\pi \alpha^{\prime}}\int d^{D}x \sqrt{-g}
e^{-2 \varphi}\Bigl( R+4(\nabla \varphi)^{2}-\frac{1}{12}H^{2}
\Bigr)+S_{\text m} \, ,
\ee
where $R$ denotes the Ricci scalar, $g$ is the determinant of the
background metric, $\varphi$ is the dilaton field, and $H$ is the field
strength of an antisymmetric tensor field. The action of the matter
sources is denoted by $S_{\text m}$.
For example, with $D=10$ this is the low energy effective action of
type II-A superstring theory.
For the purposes of this paper we will ignore the effects of
branes, since it will be the winding and momentum modes of the string that
ultimately determine the dimensionality and stability of
space-time \cite{bgc}. Here, we will ignore the effects of
fluxes \footnote{See \cite{Campos:2003ip} for inclusion of
fluxes in the scenario.}, i.e. we set $H=0$.

This action yields the following equations of motion,
\bea
\label{eomb}
& & R_{\mu}^{\; \; \nu} +2 \nabla_{\mu} \nabla^{\nu}\varphi
=8 \pi M^{-2}_{p} \; e^{2\varphi} T_{\mu}^{\; \; \nu}, \nonumber \\
& & R+4\nabla_{\kappa}\nabla^{\kappa}\varphi
-4\nabla_{\kappa}\varphi \nabla^{\kappa} \varphi=0,
\eea
where $\nabla$ is the covariant derivative.

We will work in the conformal frame with a homogeneous metric of the form
\be \label{metric}
ds^{2}=e^{2 \lambda(\eta)} \Bigl( d\eta^{2}- \delta_{i j} dx^{i} dx^{j}\Bigr)
- e^{2 \nu(\eta)} \delta_{m n} dx^{m}dx^{n},
\ee
where $(\eta,x^i)$ are the coordinates of $3+1$ space-time
and $x^m$ are the coordinates of the other six dimensions, all of which
can be
taken to be isotropic \cite{isotropization}.
The scale factors $a(\eta)$ and $b(\eta)$ are given by $\lambda \equiv
\ln(a)$ and $\nu \equiv \ln(b)$.

We consider the effect of the strings on the background through their
stress energy tensor
\be \label{stress}
T_{\mu}^{\: \: \nu} \equiv diag(\rho,-p_{i},-p_{m}),
\ee
where $\rho$ is the energy density of the strings, $p_i$ ($i=1 \ldots 3$) is
the pressure in the expanding dimensions and
$p_m$ ($m=4 \ldots 9$) is the pressure in the small dimensions (because
of our assumption of isotropy of each subspace, there is only one independent
$p_i$ and one independent $p_m$).

Strings contain winding modes, momentum modes and oscillatory modes.
However, since the energies of the oscillatory modes are independent
of the size of the dimensions, and since the winding modes and
momentum modes dominate the thermodynamic partition function at very
small and very large radii of the spatial dimensions, here we shall
neglect the oscillatory modes. In the absence of string interactions,
the contributions to the stress tensor coming from the string
winding modes and momentum
modes ($T^{\text{w}}_{\mu\nu}$ and $T^{\text{m}}_{\mu\nu}$ respectively)
are separately conserved,
\bea
T_{\mu\nu}=T^{\text{w}}_{\mu\nu}+T^{\text{m}}_{\mu\nu} \nonumber\\
\nabla^{\mu} T^{\text{w}}_{\mu\nu}=0, \;\;\;\;\;\;
\nabla^{\mu} T^{\text{m}}_{\mu\nu} \label{conservationeq}=0 \, .
\eea
The conservation equations take the form
\be \label{theconservationeq}
\rho^{\prime \: {\text w,m}} +\sum_{i=1}^9 \lambda_i^{\prime}
(\rho^{\text w,m} - p^{\text w,m}_i)=0,
\ee
where the derivatives are with respect to the conformal time $\eta$, and
where for the moment we consider 9 independent scale factors.

Expressing (\ref{eomb}) in terms of the metric (\ref{metric}) and the
stress tensor (\ref{stress}),
we find the following system of equations,
\bea \label{theset}
-3 \lpp -6 \npp +6 \lp \np -6\nps +\vpp-\lp \vp=\gn
e^{\varphi+2\lambda}\rho,\\
-\lpp+2\lps+6\lp\np+\lp\vp=-\gn e^{\varphi+2\lambda}p_{i},\\
-\npp+6\nps+2\lp\np+\vp\np=-\gn e^{\varphi+2\lambda}p_{m} \label{it},\\
-6\lpp-12\npp-24\lp\np-42\nps -6\lps -\varphi^{\prime
2}+2\vpp+8\lp\vp+12\vp\np=0. \label{set}
\eea

The explicit forms of the energy density and pressure were given in
\cite{stable}\footnote{The equations here are related to Eq. (18)
in \cite{stable} by the volume factor $V=e^{3\lambda+6\nu}$, e.g.
$\rho=\frac{E}{V}$},
\bea \label{sources1}
\rho=3 \mu N^{(3)} e^{-2 \lambda-6 \nu}+3 \mu M^{(3)} e^{-4 \lambda-6 \nu}+
6 \mu N^{(6)} e^{-3 \lambda-5 \nu}+6 \mu M^{(6)} e^{-3 \lambda-7 \nu},\\
p_i=-\mu N^{(3)} e^{-2 \lambda-6 \nu}+\mu M^{(3)} e^{-4 \lambda-6 \nu}, \\
p_m=-\mu N^{(6)} e^{-3 \lambda-5 \nu}+\mu M^{(6)} e^{-3 \lambda-7 \nu}
\label{sources1b},
\eea
where $\mu$ is a constant, $N^{(3)}$ and $M^{(3)}$ are the numbers of
winding and momentum modes in the large directions, and $N^{(6)}$ and
$M^{(6)}$ in the six small directions.

We are interested in solutions that stabilize the internal
dimensions, while allowing the three large dimensions to expand.
Such solutions were discussed in \cite{stable}, where it was shown
that the winding and momentum modes of the strings lead naturally to
stable compactifications of the internal dimensions at the self dual
radius.  This remains true as the other three dimensions grow large, which is
possible because the string gas can maintain thermal equilibrium in three dimensions and
the string winding modes are able to annihilate. Thus, we
will set $N^{(3)}=0$.
At the self dual radius, the number of winding modes is equal to the number
of momentum modes (i.e. $N^{(6)}=M^{(6)}$) and the pressure vanishes
($p_{m}=0$).

In Ref. \cite{stable}, the solutions subject to the above
conditions on the winding and momentum numbers were found numerically.
In this paper, we wish to study the stability of these solutions towards
linear perturbations in the time interval when
the internal dimensions have stabilized and
the large dimensions give power law expansion. In the following section,
we will derive the equations for the linear fluctuations. The coefficients
in these equations depend on the background solution. We will
use analytical expressions which approximate the numerically obtained
solutions of \cite{stable}. We restrict our
initial conditions so that the evolution preserves the low energy
and small string coupling assumptions ($g_s \sim e^{2 \varphi } \ll 1$).

We can approximate a typical solution of
the equations (\ref{theset}-\ref{set}) by
\bea
\lambda(\eta)= k_1 \ln(\eta) + \lambda_{0} \;\;\; {\text or} \;\;\; a(\eta) = a_{0} \eta^{k_1}, \nonumber \\
\varphi(\eta)=-k_2 \ln(\eta) + \varphi_{0},
\label{backsoln}
\eea
where the constants $k_1$, $k_2$, $\lambda_{0}$ and $\varphi_{0}$
depend on the choice of initial conditions.
We have made use of $\nu=\nu^{\prime}=\nu^{\prime \prime}=0$, $N^{(3)}=0$,
$N^{(6)}=M^{(6)}$, $p_m=0$. Note that in this limit (\ref{it}) is
trivially satisfied.
An example of a solution yielding stabilized dimensions and three dimensions
growing large corresponds to $k_1=\frac{1}{9}$ and $k_2=\frac{9}{7}$.
The numerical solution of \cite{stable} and the analytical approximation
used in this paper are compared in Fig. 1, for the above values of the
constants $k_1$ and $k_2$.

\begin{figure}[!]
\includegraphics[totalheight=6 in,keepaspectratio]{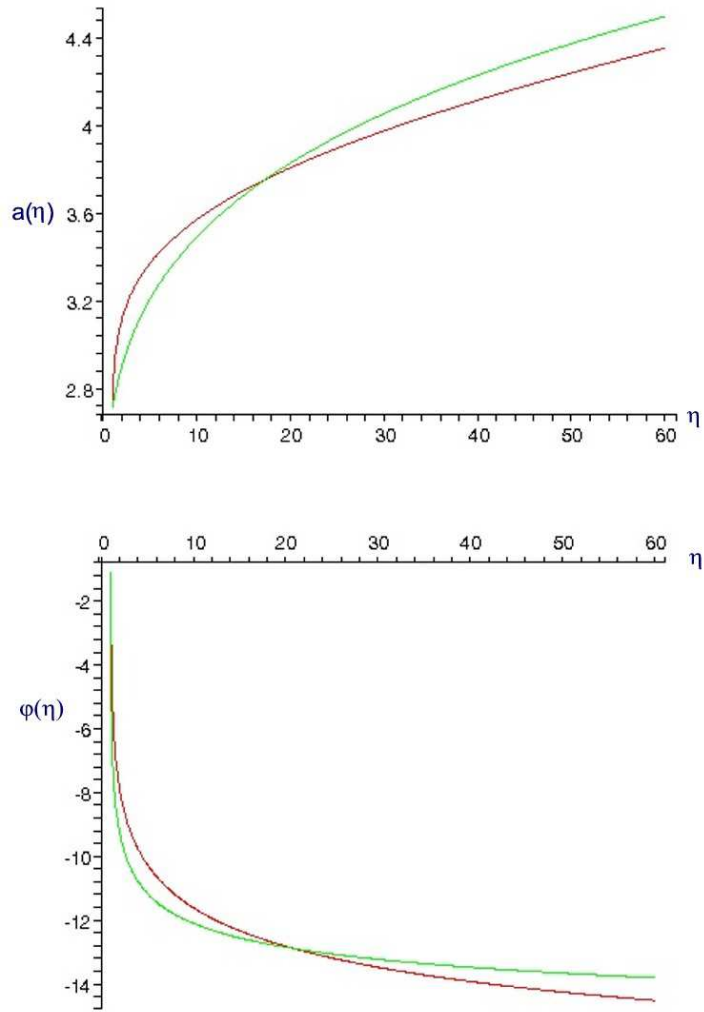}
\caption{A comparison between the numerical background solutions
obtained in \cite{stable} (red or light line) and the analytical
approximation used in this paper (green or dark line).}
\label{fig1}
\end{figure}

\section{Scalar Metric Perturbations}

In this section we consider the growth of scalar metric perturbations
(see e.g. \cite{MFB} for a comprehensive review of the theory of
cosmological perturbations) due to the presence of string inhomogeneities.
We are interested in the case where the fluctuations depend only
on the external coordinates and conformal time, not on the coordinates
of the internal dimensions. For simplicity we work in the generalized
longitudinal gauge in which the metric perturbations are only in
the diagonal metric elements \footnote{As discussed e.g. in
\cite{Dorca}, for scalar perturbations depending on all spatial coordinates
it would be inconsistent to choose the perturbed metric completely
diagonal, and one would have to add a metric coefficient to the
$dt dx^m$ terms, where $x^m$ are the coordinates of the internal
dimensions. However, as discussed in \cite{GG96}, if the fluctuations
are independent of the coordinates $x^m$, as in our case, the coefficient
can be chosen to vanish, and thus the perturbed metric is completely
diagonal.}. Thus, the metric including linear fluctuations is given by
\be \label{pertmetric}
ds^{2}=e^{2 \lambda(\eta)} \Bigl( (1+2\phi) d\eta^{2}- (1-2\psi)\delta_{i j}
dx^{i} dx^{j}\Bigr) -e^{2 \nu(\eta)}(1-2\xi) \delta_{m n}
dx^{m}dx^{n}.
\ee
The dilaton $\varphi$ also fluctuates about its background
value $\varphi_0$. The dilaton fluctuation $\chi$ is determined by
\be
\varphi = \varphi_0 + \delta \varphi, \;\;\;\;\;\; \chi \equiv
\delta \varphi.
\ee
In the above, the fluctuating fields $\chi, \phi, \psi$ and $\xi$ are
functions of the external coordinates $x^i$ and time, i.e.
\be
\chi=\chi(\eta,x^i), \: \: \: \: \phi=\phi(\eta,x^i), \: \: \: \:  \psi=\psi(\eta,x^i),\: \: \: \:
\xi=\xi(\eta,x^i).
\ee

The perturbations of the matter energy momentum tensor result
from over-densities and under-densities in the number of strings.
From (\ref{sources1})-(\ref{sources1b}) and noting
that we are interested in the case when $N^{(3)}=0$ and
$M^{(6)}=N^{(6)}$ we find,
\bea &\delta \rho=\delta \rho_{{\text w}} + \delta \rho_{{\text
m}},\\ &\delta \rho_{{\text w}}=30 \mu N \xi e^{-3\lambda}+18 \mu
N \psi e^{-3\lambda}+6 \delta N^{(6)}e^{-3\lambda},
\\ &\delta \rho_{{\text m}}=42 \mu N \xi e^{-3\lambda}+18 \mu M \xi e^{-4\lambda}+
18 \mu N \psi e^{-3\lambda}+12 \mu M \psi e^{-4\lambda}+6 \mu
\delta M^{(6)}e^{-3\lambda}+\\&+3 \mu \delta M e^{-4\lambda} ,
\\ &\delta p_{\lambda}=6 \mu M \xi e^{-4 \lambda}+4 \mu M \psi e^{-4 \lambda}+
\mu \delta M e^{-4 \lambda},\\
&\delta p_{\nu}=2 \mu N \xi e^{-3 \lambda}-\mu \delta N^{(6)}
e^{-3 \lambda}+\mu \delta M^{(6)} e^{-3 \lambda}, \eea
where we define\footnote{Notice that we must be careful to
distinguish between the perturbed quantities $\delta N^{(6)}$ and
$\delta M^{(6)}$.} $N \equiv N^{(6)}=M^{(6)}$ and $M \equiv
M^{(3)}$. The fluctuations $\delta N^{(6)}$, $\delta M^{(6)}$, and
$\delta M$ are taken as functions of both conformal time and the
external space, e.g. $\delta N^{(6)}=\delta N(\eta, x^i)$.

It follows from (\ref{theconservationeq}) that the perturbed sources
obey modified conservation equations for both the winding and momentum modes,
\be \label{pce}
\delta \rho^{\prime \: {\text w,m}} +\sum_{i=1}^9 \lambda_{i}^{\prime} (\delta \rho^{\text w,m}
- \delta p^{\text w,m}_i)+\sum_{i=1}^9 \delta \lambda_i^{\prime} (\rho^{\text w,m}
- p^{\text w,m}_i)=0,
\ee
where $\delta \lambda=a^{-1} \delta a =-\psi$ and $\delta
\nu=b^{-1} \delta b=-\xi$ are spatial variations.

We rewrite (\ref{eomb}) to take the more familiar form of the
Einstein and dilaton equations, namely

\bea
R_{\mu}^{\nu}-\frac{1}{2} \delta_{\mu}^{\nu} R=e^{2\varphi} T_{\mu}^{\nu}-2 g^{\alpha \nu}
\nabla_{\mu} \nabla_{\alpha} \varphi + 2 \delta_{\mu}^{\nu} \Bigl( g^{\mu \nu} \nabla_{\mu} \nabla_{\nu} \varphi
-g^{\mu \nu} \partial_{\mu} \varphi \partial_{\nu} \varphi
\Bigr),  \nonumber \\
g^{\mu \nu} \partial_{\mu} \varphi \partial_{\nu} \varphi
-\frac{1}{2} g^{\mu \nu} \nabla_{\mu} \nabla_{\nu}\varphi=\frac{1}{4}
e^{2 \varphi} T^{\mu}_{\mu},
\eea
where we invoke Planckian units (i.e. $\gn= 1$).
Plugging the perturbed metric (\ref{pertmetric}) and dilaton into
these equations, making use of the background equations of motion,
and linearizing the equations about the background (i.e. keeping
only terms linear in the fluctuations) yields the following set of
equations:
\bea \label{eq111}
{\vec \nabla}^2 \psi +3 {\vec \nabla}^2\xi -9 \ch \xp -3\ch \pp -3
\ch^2 \phi =\frac{1}{2} e^{2\varphi + 2\lambda}\Bigl( 2\chi T^0_0 +\delta T^0_0 \Bigr)
-6 \ch \phi \vp \nonumber \\-3 \pp \vp-6\xp \vp -{\vec \nabla}^2 \chi +3 \ch
\cp +2 \phi \varphi^{\prime 2}-2\cp \vp,
\eea
\be
\partial_{i}\pp+3\partial_{i}\xp+\ch\partial_{i}\phi-3\ch\partial_{i}\xi=
\frac{1}{2} e^{2\varphi+2\lambda} \delta T_{0}^{i}+\partial_{i}\phi
\vp-\partial_{i}\cp+\ch\partial_{i}\chi,
\ee
\be \label{usefull}
\partial_{i}\partial_{j} \Bigl( \phi-\psi-6\xi-2\chi \Bigr)=0 \;\;\;\;\;\; i \neq j,
\ee
\begin{widetext}
\bea
\Bigl( \partial_{i}^{2}-\vec{\nabla}^{2} \Bigr)\Bigl( \phi-\psi-6\xi
\Bigr)-2\ppp-6\xpp-
4\ch\pp -6\ch\xp-2\ch{^2}\phi-4\ch^{\prime}\phi-2\fp\ch
\nonumber \\ =e^{2 \varphi+2\lambda} \Bigl( 2\chi T^{i}_{i}+\delta T^{i}_{i} \Bigr)
+2\partial_{i}^{2}\chi-4\phi\vpp-2\fp\vp  -4\ch\phi\vp-4\pp\fp-12\xp\vp\nonumber\\
+2\cpp-2\vec{\nabla}^{2}\chi+2\ch\cp+4\phi\varphi^{\prime
2}-4\cp\vp,
\eea
\end{widetext}
\bea
-{\vec \nabla}^2 \phi +5 {\vec \nabla}^2 \xi -5 \xpp +2 {\vec \nabla}^2
\psi-3 \ppp-10 \ch \xp -3 \fp \ch -9 \ch \pp -6 \ch^2 \phi -6
\ch^{\prime} \phi \nonumber \\ =
 e^{2\varphi+2\lambda} \Bigl( 2 \chi T^m_m + \delta T^m_m \Bigr)
 -4 \phi \vpp -2 \fp \vp -8 \ch \phi \vp -6 \pp \vp -10 \xp \vp  +2
 \cpp \nonumber \\-2{\vec \nabla}^2 \chi +4 \ch \cp +4 \phi \varphi^{\prime 2}
 -4 \cp \fp,
\eea
\bea \label{eq113}
-2 \phi \varphi^{\prime 2}+2 \vp \cp+\phi \vpp -6 \psi \ch
\vp+\frac{1}{2} \fp \vp -2 \ch \phi \vp - \frac{3}{2} \pp \vp -3
\xp \vp -\frac{1}{2}\cpp \nonumber \\ +\frac{1}{2}{\vec \nabla}^2
\chi -\ch \cp
=\frac{1}{4}e^{2\varphi + 2\lambda} \Bigl( 2\chi T + \delta T
\Bigr),
\eea
where $T\equiv T^{\mu}_{\mu}$ is the trace of the stress tensor and $\vec{\nabla}^2 \equiv \partial_x^2+\partial_y^2+\partial_z^2$ is the spatial
Laplacian.
The modified conservation equations (\ref{pce}) take the form
\bea \label{yep1b}
&\frac{d }{d \eta}\Bigl( \delta N^{(6)} \Bigr)=7 N \xp,\\
&42 \mu N \xp-72 \mu M \xi \lp e^{- \lambda}-48 \mu M \psi \lp e^{- \lambda}
+12 \mu M \pp e^{- \lambda}+6 \mu \frac{d }{d \eta}\Bigl( \delta M^{(6)}\Bigr)\nonumber \\&
-12 \mu \delta M \lp e^{- \lambda} +3 \mu \frac{d }{d \eta}\Bigl( \delta M \Bigr) e^{-
\lambda}=0.
\label{yep1} \eea
These equations give us the evolution of the metric perturbations
$\phi$, $\psi$, and $\xi$ in terms of the matter perturbations
$\chi$, $\delta \rho$, and $\delta p_i$.  At first glance, it may
appear that the above system is over-determined since we have
eight equations for seven unknowns.  However, as is the case in
standard cosmology, the conservation equations are not independent
of the Einstein equations. Thus, we can choose to keep only one
of the modified conservation equations and our system will be
consistent.
\section{Ultraviolet Modes}
\begin{figure}[!]
\includegraphics[totalheight=3.5 in,keepaspectratio]{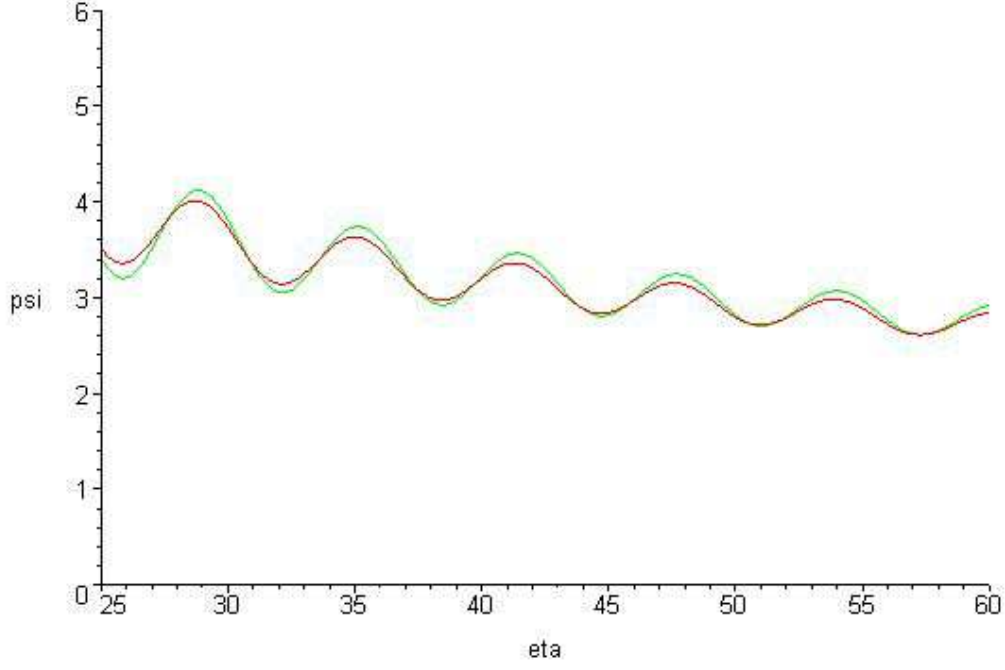}
\caption{A comparison between the numerical value of the
external metric perturbation $\psi$ (red or dark line) and the analytical
approximation found in this paper (green or light line).}
\label{fig2}
\end{figure}
\begin{figure}[!]
\includegraphics[totalheight=3.5 in,keepaspectratio]{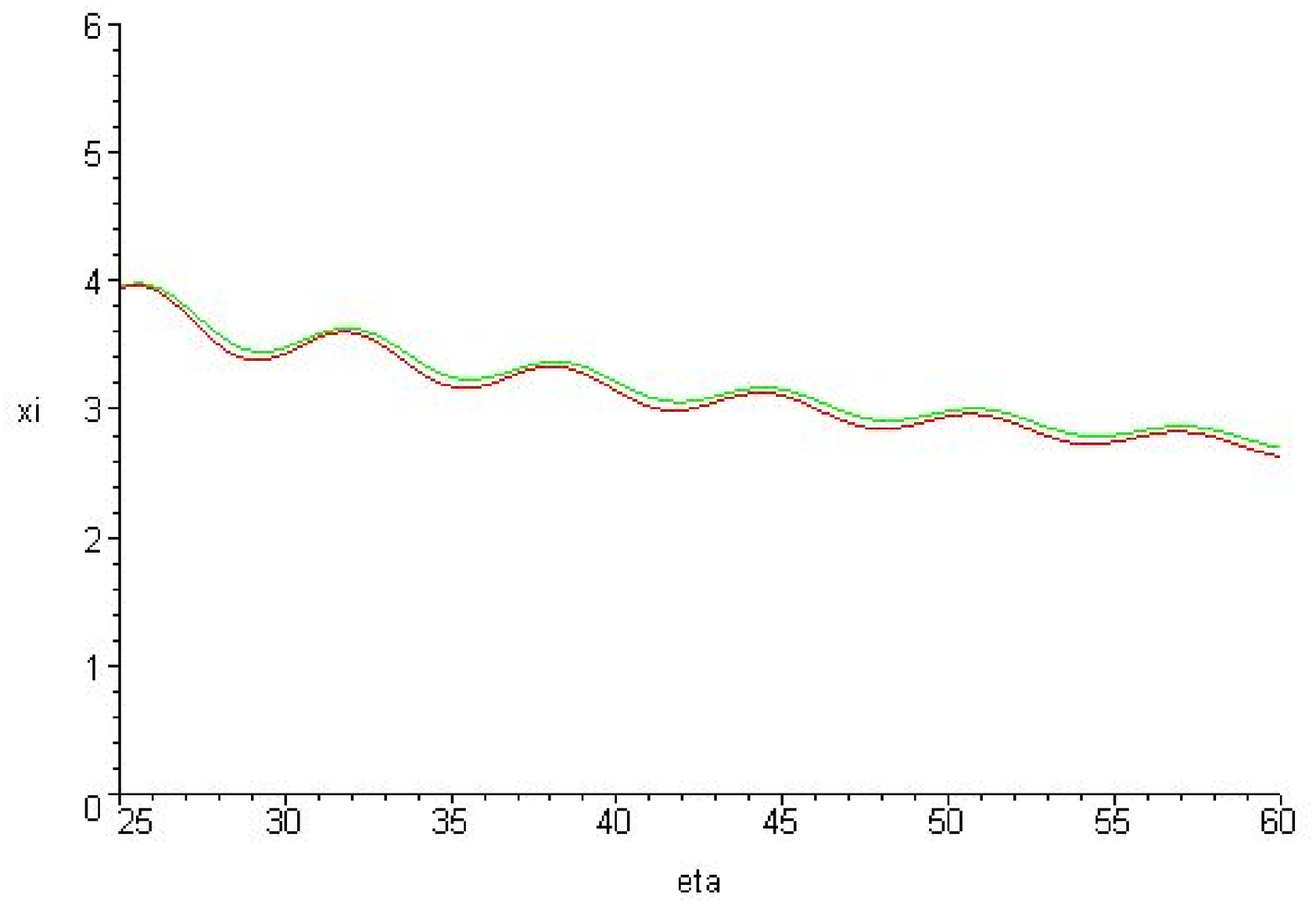}
\caption{A comparison between the numerical value of the
internal metric perturbation $\xi$ (red or dark line) and the analytical
approximation found in this paper (green or light line).}
\label{fig3}
\end{figure}
\begin{figure}[!]
\includegraphics[totalheight=3.5 in,keepaspectratio]{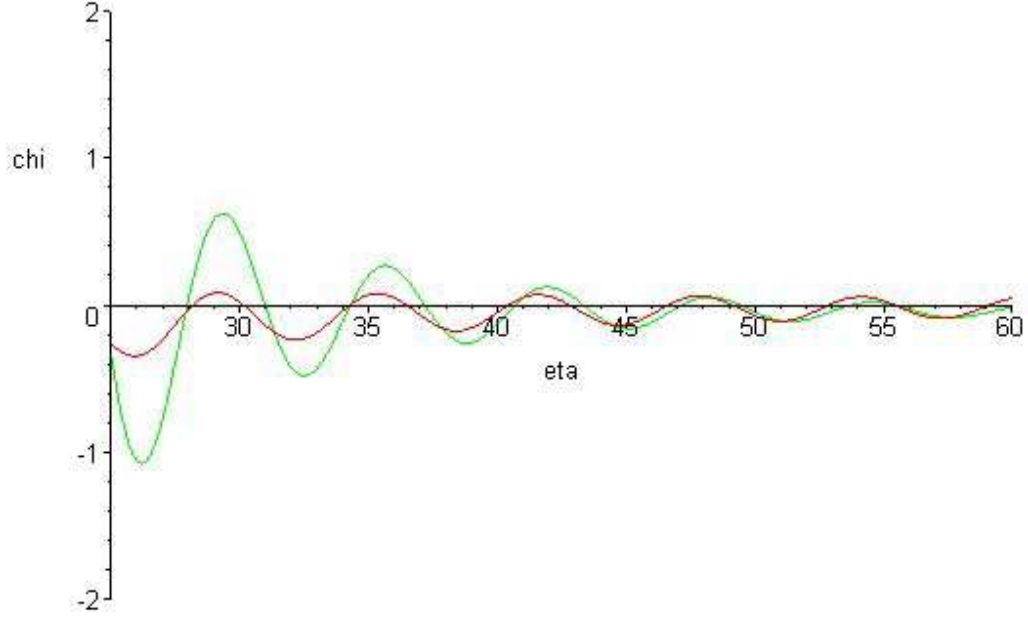}
\caption{A comparison between the numerical value of the dilaton
perturbation $\chi$ (red or dark line) and the analytical
approximation found in this paper (green or light line).}
\label{fig4}
\end{figure}
We now want to solve the equations (\ref{eq111})-(\ref{eq113}) in the
limit of small wavelength (or high energy).  We can simplify the
analysis
by working in terms of the Fourier modes, e.g. $$
\psi(\eta,\vec{x})=\sum_k \psi_k(\eta) e^{i \vec{k} \cdot \vec{x}},
\; \; \; \delta N(\eta,\vec{x})=\sum_k \delta N_k(\eta) e^{i \vec{k} \cdot
\vec{x}}, \; \; \; {\text etc...}.$$
Note that in the remainder of this paper it will be understood that
when we speak of perturbed quantities
we are referring to the time dependent Fourier modes, e.g. $\psi
\equiv \psi_k(\eta)$.

Using (\ref{usefull})
to eliminate the scalar metric
perturbation $\phi$,
the equations (\ref{eq111})-(\ref{eq113}) in Fourier space take
the form
\be \label{sw1}
\frac{61}{7 \eta} \xp +
\frac{61}{21 \eta} \cp + \frac{88}{21 \eta} \pp +
\Biggl( \frac{338}{189 \eta^2}+\vec{k}^2 \Biggr) \chi
+\Biggl( \frac{338}{63 \eta^2}+3\vec{k}^2 \Biggr) \xi
+\Biggl( \frac{169}{189 \eta^2}+ \vec{k}^2 \Biggr) \psi=0,
\ee
\be \label{sw2}
\pp+3 \xp +\cp +\frac{88}{63 \eta} \psi +\frac{169}{21 \eta} \xi + \frac{169}{63 \eta}
\chi=0
\ee
\be \label{sw3}
2 \ppp + 6 \xpp +2 \cpp +\frac{176}{21 \eta}\pp +\frac{230}{7 \eta} \xp
+\frac{230}{21 \eta} \cp +\frac{6434}{3969 \eta^2} \psi +\frac{12868}{1323 \eta^2} \xi +\frac{12868}{3969 \eta^2}
\chi=0,
\ee
\be \label{sw4}
3 \ppp + 5\xpp + 2 \cpp + \frac{244}{21 \eta} \pp +\frac{1978}{63 \eta} \xp +\frac{718}{63
\eta}\cp+\Biggl( \frac{2672}{1323 \eta^2}+\vec{k}^2 \Biggr) \psi
+\Biggl( \frac{5344}{441 \eta^2}-\vec{k}^2 \Biggr) \xi +\frac{5344}{1323 \eta^2}
\chi=0,
\ee
\be \label{sw5}
\frac{1}{2}\cpp-\frac{9}{7 \eta} \pp +\frac{250}{63 \eta} \cp +\frac{43}{49 \eta^2}
\psi+\frac{510}{49 \eta^2} \xi
+\Biggl( \frac{170}{49 \eta^2}+ \frac{1}{2} \vec{k}^2 \Biggr) \chi=0.
\ee
To obtain these equations we have made use of the background solution (\ref{backsoln}) and
dropped all but the leading order terms since we
are interested in the late time ($\eta >>1$) and
small wavelength ($k >> 1$) behavior \footnote{For the interested reader, the full equations are
presented in the Appendix.}.
In particular, we see that as in the long wavelength case, the source terms
$\delta N$, $\delta M_3$, and $\delta M_6$ are negligible at late
times. This is because the string matter sources are sub-leading in the evolution
equations and
instability is primarily sourced by the dilaton perturbation.
This result is crucial to our outcome and is discussed in detail
in the Appendix.

We also notice that (\ref{sw1}) and (\ref{sw2}) are only
first order in time derivatives and can be taken as constraints on the initial conditions.
This leaves us with the equations of
motion (\ref{sw3}), (\ref{sw4}), and (\ref{sw5}).
These equations can be put in a more tractable form by introducing
the two fields $\Delta$ and $\Theta$,
\be
\psi=\Delta+\Theta \;\;\;\;\;
\xi=\Delta- \frac{1}{3}\Theta.
\ee
The equations can then be written as
\bea \label{s1}
\frac{1}{2}\cpp +\frac{250}{63 \eta} \cp +\Biggl( \frac{170}{49 \eta^2} + \frac{1}{2}
\vec{k}^2 \Biggr)
\chi=\frac{9}{7 \eta} \Dp+\frac{9}{7 \eta} \Tp -\frac{79}{7 \eta^2} \Delta +\frac{127}{49 \eta^2}
\Theta\\ \label{s2}
\frac{4}{3}\Tpp+\frac{704}{189 \eta} \Tp +\Biggl( \frac{4}{3}
\vec{k}^2-\frac{226}{567 \eta^2}\Biggr)
\Theta=-\frac{16}{9 \eta}\Dp - \frac{4}{9 \eta} \cp-\frac{226}{81
\eta^2}\Delta -\frac{452}{567 \eta^2}\chi\\
8\Dpp+\frac{866}{21 \eta}\Dp+\frac{6434}{567
\eta^2}\Delta=-2\cpp-\frac{230}{21 \eta}\cp +\frac{18}{7 \eta}\Tp +\frac{6434}{3969 \eta^2}\Theta
-\frac{12868}{3969 \eta^2}\chi \label{s3}
\eea
where we have written the system as to isolate the second order
derivative terms in $\Theta$ and $\Delta$ and again dropped terms that are negligible given $k, \eta >>1$.
We first solve (\ref{s1}) for $\chi$,
neglecting the right side of the equation and treating it as a
negligible source term.  This perturbative approach will only be
justified if, after solving for $\Delta$ and $\Theta$ in the remaining
equations,
we return to
(\ref{s1}) to make sure these terms remain negligible.
Proceeding in this way we find that to first order $\chi$ is
given by,
\be
\chi_0= c_1(k) \frac{J_{a}(k\eta)}{(k \eta)^{a}}+c_2(k)
\frac{Y_{a}(k\eta)}{(k \eta)^{a}},
\ee
where the $c_i(k)$ are arbitrary constants, $a=\frac{147}{126}$, $J_{a}$ is a Bessel Function of the
first kind, and $Y_{a}$ is a Bessel Function of the second kind.
We proceed by solving (\ref{s2}) for $\Theta$ using $\chi=\chi_0$
and again neglecting the terms that depend on $\Delta$.  Thus, we
wish to solve the equation
\be
\frac{4}{3}\Tpp+\frac{704}{189 \eta}\Tp+\Biggl( \frac{4}{3}\vec{k}^2
-\frac{226}{567 \eta^2} \Biggr)
\Theta=S_{\Theta}(\eta),
\ee
where
\bea
S_{\Theta}(\eta)=
 c_3(k) \frac{J_{b}(k \eta)}{(k \eta)^{c}}
+c_4(k) \frac{J_{d}(k \eta)}{(k \eta)^{e}}
+c_5(k) \frac{J_{d}(k \eta)}{(k \eta)^{f}}
+c_6(k) \frac{J_{b}(k \eta)}{(k \eta)^{g}}
+c_7(k) \frac{Y_{d}(k \eta)}{(k \eta)^{d}}\nonumber \\
+c_8(k) \frac{Y_{b}(k \eta)}{(k \eta)^{c}}
+c_9(k) \frac{Y_{d}(k \eta)}{(k \eta)^{f}}
+c_{10}(k) \frac{Y_{b}(k \eta)}{(k \eta)^{g}},
\eea
where again the $c_i$'s are arbitrary constants and $b=\frac{59}{126}$, $c=\frac{815}{126}$,
$d=\frac{185}{126}$, $e=\frac{941}{126}$, $f=\frac{689}{126}$, and $g=\frac{563}{126}$.
The solution is given by
\be
\Theta_0=c_{11}(k) \frac{J_{h}(k\eta)}{(k \eta)^{h}}+c_{12}(k)
\frac{Y_{h}(k\eta)}{(k \eta)^{h}}+c_{13}(k)\int
G_{\Theta}(\eta-\eta^{\prime}) S_{\Theta}(\eta^{\prime})
d\eta^{\prime},
\ee
where $G_{\Theta}$ is the Green's function
\be
G_{\Theta}(\eta,\eta^{\prime})=\frac{3 \pi}{8 \eta^{h}\eta^{\prime -i}}\Biggl( {Y_{h}(k \eta)J_{h}(k
\eta^{\prime})}-{J_{h}(k \eta)Y_{h}(k \eta^{\prime})}\Biggr),
\ee
with $h=\frac{113}{126}$ and $i=\frac{239}{126}$.
This is valid for $\eta > \eta^{\prime}$ and $G_{\Theta}$ vanishes otherwise.
On evaluating the source integral we find that the leading behavior of
$\Theta$ is given by the homogeneous part of the solution, i.e.
\be
\Theta_0 \sim c_{11}(k) \frac{J_{h}(k\eta)}{(k \eta)^{h}}+c_{12}(k)
\frac{Y_{h}(k\eta)}{(k \eta)^{h}}
\ee
Using this result in (\ref{s3}) we finally find an equation for $\Delta$
\be
8\Dpp+\frac{866}{21 \eta}\Dp+\frac{6434}{567 \eta^2}
\Delta=S_{\Delta}(\eta),
\ee
where
\bea
S_{\Delta}(\eta)=
 c_{14}(k) \frac{J_{h}(k \eta)}{(k \eta)^{j}}
+c_{15}(k) \frac{J_{b}(k \eta)}{(k \eta)^{g}}
+c_{16}(k) \frac{J_{d}(k \eta)}{(k \eta)^{l}}
+c_{17}(k) \frac{J_{i}(k \eta)}{(k \eta)^{i}}
+c_{18}(k) \frac{J_{b}(k \eta)}{(k \eta)^{c}}\nonumber \\
+c_{19}(k) \frac{J_{d}(k \eta)}{(k \eta)^{f}}
+c_{20}(k) \frac{J_{d}(k \eta)}{(k \eta)^{e}}
+c_{21}(k) \frac{Y_{h}(k \eta)}{(k \eta)^{j}}
+c_{22}(k) \frac{Y_{b}(k \eta)}{(k \eta)^{g}}
+c_{23}(k) \frac{Y_{d}(k \eta)}{(k \eta)^{l}}\nonumber \\
+c_{24}(k) \frac{Y_{i}(k \eta)}{(k \eta)^{i}}
+c_{25}(k) \frac{Y_{b}(k \eta)}{(k \eta)^{c}}
+c_{26}(k) \frac{Y_{d}(k \eta)}{(k \eta)^{f}}
+c_{27}(k) \frac{Y_{d}(k \eta)}{(k \eta)^{e}},\nonumber \\
\eea
where $j=\frac{365}{126}$ and $l=\frac{437}{126}$.  The solution
is given by
\be
\Delta(\eta)=\frac{c_{28}(k)}{(k \eta)^{m-n}}+ \frac{c_{29}(k)}{(k \eta)^{m+n}}
+c_{30}(k)\int G_{\Delta}(\eta-\eta^{\prime}) S_{\Delta}(\eta^{\prime})
d\eta^{\prime},
\ee
where $m=\frac{349}{168}$, $n=\frac{\sqrt{735905}}{504}$, and $G_{\Delta}$ is given by
\be
G_{\Delta}(\eta,\eta^{\prime})=\frac{\alpha \eta^{\prime p}}{\eta^m}
\Biggl[ \Biggl(\frac{\eta}{\eta^{\prime}}\Biggr)^n-\Biggl(\frac{\eta^{\prime}}{\eta}\Biggr)^n \Biggr],
\ee
with $p=\frac{517}{168}$.  The Green's function,  $G_{\Delta}$,
is valid for $\eta > \eta^{\prime}$ and vanishes otherwise.
Using this result for $\Delta$ and $\Theta_0$ one can check that we were justified
in neglected the terms in both (\ref{s1}) and (\ref{s2}).  That is, these terms do not significantly
change the evolution.  Thus, we have found that there are no
growing exponential instabilities.  In fact, we find that the
behavior of the perturbations is that of a decaying oscillator.

As another check of our approximation, we can compare our analytic
solution with a numerical treatment.  By approximating $\Delta$
and $\Theta_0$ as we have discussed, i.e. ignoring the source
terms, we find the following approximate form for the
perturbations
\bea \label{solutions}
\psi \sim \eta^{-m+n}+\eta^{-m-n}+\frac{\cos(k \eta+\delta_1)}{\eta^q},\nonumber\\
\xi \sim \eta^{-m+n}+\eta^{-m-n}-\frac{1}{3}\frac{\cos(k
\eta+\delta_2)}{\eta^q},\nonumber\\
\chi \sim \frac{\cos(k \eta+\delta_3)}{\eta^r},
\eea
where we have used the asymptotic form of the Bessel functions, $q=\frac{88}{63}$,
$r=\frac{250}{63}$ and the $\delta_i$ represent time-independent phases.
In Fig. (\ref{fig2}), Fig. (\ref{fig3}), and Fig. (\ref{fig4}) we compare
these approximate solutions to the numerical solution of the full
equations (\ref{s1})-(\ref{s3}).  We find agreement at late times (large $\eta$)
giving us a second check that our approximations were warranted.
Thus, we conclude that the small wavelength or ultraviolet
perturbations are well behaved in the linear regime.

\section{Conclusions}
We have extended the analysis of perturbations in BGC to include
the UV modes.  We have derived the evolution equations for the fluctuations
at small wavelengths and at late times.  We then solved these equations using a perturbative
approach, which we were able to check both analytically and numerically.
We find a novel behavior for the perturbations, in that string matter sources are
negligible compared with the dilaton perturbation and the
resulting behavior is that of a decaying oscillator.  This has
interesting consequences in regards to the worry of black hole formation and
the usual worrisome behavior of Kaluza-Klein massive states on the
background.  We have concluded that at the linear level and in the gas approximation
these types of string
matter sources will have a negligible effect.
Moreover, we find that the predictions of BGC remain robust under the
consideration of both long and short wavelength perturbations.  In
particular, the prediction that $3+1$ dimensions will grow large while
$6$ dimensions remain stabilized
around the self dual radius remains intact.

Although these results are promising for BGC there is still much
to be done. A more complete treatment of the perturbations would need to take
into consideration the non-linear behavior.  It would also be
interesting to test the string gas approach itself.  That is, how
does one go from the consideration of the effects of individual
strings to the known predictions of BGC? Finally, it is an important
consideration to reexamine these perturbations in the presence of a frozen dilaton. We
know that at very late times in the cosmological evolution the dilaton
most likely acquired a mass.  Since the dilaton perturbation played such a
vital role in this analysis it could be expected that the results would change dramatically
in the massive dilaton case. However, if the perturbations do remain well
behaved in this case, it would also be of interest to see if BGC could give rise to a method of structure
formation or a unique signature to be observed in the Cosmic
Microwave Background.  We leave these questions and concerns to future work.

\begin{acknowledgments}

SW would like to thank Robert Brandenberger and Sera Cremonini
for many critical comments and suggestions throughout the duration of this project.
SW would also like to thank the University of North Carolina at
Wilmington for their generous hospitality.
This research was supported in part by the NASA Graduate Student Research Program.

\end{acknowledgments}

\appendix
\section{Perturbation Equations for UV Modes}
In this appendix we will examine in more detail the arguments that
led to the string matter source terms being dropped from
(\ref{sw1})-(\ref{sw5}).
We begin by introducing the Fourier modes,
\bea
\psi(\eta,\vec{x})=\sum_k \psi_k(\eta) e^{i \vec{k} \cdot \vec{x}}
, \;\;\;\;\; \psi_k(\eta) \equiv \psi,\\
\xi(\eta,\vec{x})=\sum_k \xi_k(\eta) e^{i \vec{k} \cdot \vec{x}}
, \;\;\;\;\; \xi_k(\eta) \equiv \xi,\\
\chi(\eta,\vec{x})=\sum_k \chi_k(\eta) e^{i \vec{k} \cdot \vec{x}}
, \;\;\;\;\; \chi_k(\eta) \equiv \chi,\\
\delta M(\eta,\vec{x})=\sum_k \delta M_k(\eta) e^{i \vec{k} \cdot
\vec{x}}, \;\;\;\;\; \delta M_k(\eta) \equiv \delta \tilde{M},\\
\delta N^{(6)}(\eta,\vec{x})=\sum_k \delta N^{(6)}_k(\eta) e^{i \vec{k} \cdot
\vec{x}}, \;\;\;\;\; \delta N^{(6)}_k(\eta) \equiv \delta \tilde{N}^{(6)},\\
\delta M^{(6)}(\eta,\vec{x})=\sum_k \delta M^{(6)}_k(\eta) e^{i \vec{k} \cdot
\vec{x}}, \;\;\;\;\; \delta M^{(6)}_k(\eta) \equiv \delta \tilde{M}^{(6)},\\
\delta{T}^{i0}(\eta,\vec{x})=\sum_k \delta {T}^{i0}_k(\eta) e^{i \vec{k} \cdot
\vec{x}}, \;\;\;\;\; \delta {T}^{i0}_k(\eta) \equiv \delta \tilde{T}^{i0} .
\eea

Note that in the remainder of this paper it will be understood that
when we speak of perturbed quantities
we are referring to the time dependent Fourier modes, e.g. $\psi
\equiv \psi_k(\eta)$. Given these modes, the equations (\ref{eq111})-(\ref{eq113})
now become
\bea \label{asw1}
\Biggl(\frac{61}{21 \eta}+\frac{324}{49 \eta^2} \Biggr) \cp +
\Biggl( \frac{88}{21 \eta}+\frac{162}{49 \eta^2} \Biggr)
\pp+
\Biggl( \frac{12 N}{ \eta^{\frac{169}{63}}} +\frac{3 M}{ \eta^{\frac{176}{63}}}
+\vec{k}^2 +\frac{338}{189 \eta^2} \Biggr) \chi+\nonumber\\+\Biggl(\frac{972}{49 \eta^2}+\frac{61}{7 \eta} \Biggr) \xp
+\Biggl( 3\vec{k}^2+\frac{36 N}{ \eta^{\frac{169}{63}}}
+\frac{9 M}{ \eta^{\frac{176}{63}}}+\frac{338}{63 \eta^{2}}  \Biggr)
\xi+
3\frac{ \delta N}{ \eta^{\frac{169}{63}}}
+\nonumber\\+ \Biggl( \frac{6 M}{ \eta^{\frac{176}{63}}} +\frac{169}{189 \eta^{2}} +\frac{18 N}{ \eta^{\frac{169}{63}}}
+\vec{k}^2  \Biggr) \psi +\frac{3}{2}\frac{ \delta M}{
\eta^{\frac{176}{63}}}+3\frac{ \delta M^{(6)}}{ \eta^{\frac{169}{63}}}
=0,\;\;\;\;
\eea
\be \label{asw2}
\pp+3 \xp +\cp +\frac{88}{63 \eta} \psi +\frac{169}{21 \eta} \xi + \frac{169}{63 \eta}
\chi+\frac{\delta \tilde{T}_{i0}}{\eta^{\frac{148}{63}}}=0
\ee
\bea \label{asw3}
2 \ppp + 6 \xpp +2 \cpp +\frac{176}{21 \eta}\pp +\frac{230}{7 \eta} \xp
+\frac{230}{21 \eta} \cp
+\Biggl( \frac{6434}{3969 \eta^2} + \frac{4 M}{ \eta^{\frac{176}{63}}} \Biggr)\psi
+\nonumber\\+\Biggl( \frac{12868}{1323 \eta^2}+ \frac{6 M}{ \eta^{\frac{176}{63}}} \Biggr)\xi
+\Biggl( \frac{12868}{3969 \eta^2}+ \frac{2 M}{ \eta^{\frac{176}{63}}} \Biggr)
\chi+\frac{\delta M}{ \eta^{\frac{176}{63}}}=0,
\eea
\bea \label{asw4}
3 \ppp + 5\xpp + 2 \cpp + \frac{244}{21 \eta} \pp +\frac{1978}{63 \eta} \xp
+\frac{718}{63 \eta}\cp+\Biggl( \frac{2672}{1323 \eta^2}
+\vec{k}^2 \Biggr)\psi+\nonumber\\
+\Biggl(\frac{5344}{441 \eta^2}+\frac{2 N}{ \eta^{\frac{169}{63}}}-\vec{k}^2 \Biggr)\xi
+\frac{5344}{1323 \eta^2}\chi + \frac{\delta M^{(6)}}{ \eta^{\frac{169}{63}}}
+ \frac{\delta N^{(6)}}{ \eta^{\frac{169}{63}}}=0,
\eea
\bea \label{asw5}
\frac{1}{2}\cpp-\frac{9}{7 \eta} \pp +\frac{250}{63 \eta} \cp
+\Biggl( \frac{18 N}{2 \eta^{\frac{169}{63}}}+\frac{43}{49 \eta^2} \Biggr)\psi
+\Biggl( \frac{30 N}{2 \eta^{\frac{169}{63}}}+\frac{510}{49 \eta^2}
\Biggr)\xi
+\nonumber\\+\Biggl( \frac{6 N}{\eta^{\frac{169}{63}}}+\frac{170}{49 \eta^{2}}+\frac{1}{2} \vec{k}^2  \Biggr)\chi
+\frac{3 \delta N^{(6)}}{ \eta^{\frac{169}{63}}}=0.\;\;\;\;\;
\eea
From these equations we see that for late times ($\eta>>1$) and
small wavelengths ($k>>1$) a number of terms can be neglected
and we arrive at equations (\ref{sw1})-(\ref{sw5}).  In
particular, notice that the string matter perturbations $\delta M$, $\delta
N^{(6)}$, $\delta M^{(6)}$ appear to be negligible compared to the other terms.
This means that the dilaton perturbation $\delta \varphi = \chi$
is the most important source of the scalar metric perturbation.
Of course, depending on the time dependence of the string
perturbations it could be that these terms are not negligible.
We can test our assumption in the following way.  In Section 4, by
neglecting these (and other terms of explicit higher order) we
found the approximate solutions (\ref{solutions}).
\bea
\psi \sim \eta^{-m+n}+\eta^{-m-n}+\frac{\cos(k \eta+\delta_1)}{\eta^q},\nonumber\\
\xi \sim \eta^{-m+n}+\eta^{-m-n}-\frac{1}{3}\frac{\cos(k
\eta+\delta_2)}{\eta^q},\nonumber\\
\chi \sim \frac{\cos(k \eta+\delta_3)}{\eta^r},
\eea
We must now
plug these quantities back into the full equations (\ref{asw1})-(\ref{asw5}) and check
that the negligible quantities remain negligible.  However, in the
case of the string matter perturbations it turns out that we can perform another
check.  For example, in the case of the perturbation $\delta
N^{(6)}$ we can use the conservation equation (\ref{yep1b}) to find
\be
\delta
N^{(6)}=7 N \xi + constant.
\nonumber
\ee
By plugging this into (\ref{asw1})-(\ref{asw5}) we see that the
term is indeed negligible compared to the other terms.  Similarly,
this can be shown for the other two matter perturbations using the
conservation equation (\ref{yep1}) and the constraint equation (\ref{asw1}).
Thus, we have demonstrated that the matter perturbation is
negligible and the dilaton perturbation is the primary source of
the fluctuations.

\end{document}